\documentclass[]{aa}

\usepackage{graphicx,amssymb,psfig,natbib}

\newcommand{\AaA}{A\&A}

\newcommand{\ApJS}{Astrophysical Journal Supplement Series}
\newcommand{\ApJ}{ApJ}
\newcommand{\MNRAS}{MNRAS}
\newcommand{\RvMP}{Reviews of Modern Physics}

\newcommand{\Sci}{Science}

\begin{document}

\title{Some empirical estimates of the H$_2$ formation rate in photon-dominated regions}

\author{E. Habart$^1$, F. Boulanger$^2$, L. Verstraete$^2$, C.M. Walmsley$^1$, G. Pineau des For\^ets$^2$}

\institute{$^1$ Osservatorio Astrofisico di Arcetri, INAF, Largo E. Fermi 5, I-50125 Firenze, Italy\\
$^2$ Institut d'Astrophysique Spatiale, Universit\'e Paris-Sud, 91405 Orsay, France} 
 
\date{Received 2002 December 20 / accepted 2003 july 25} 

\abstract{
We combine recent ISO observations of the vibrational ground state lines of H$_2$
towards Photon-Dominated Regions (PDRs) with observations of vibrationally 
excited states made with ground-based telescopes in order to constrain the
formation rate of H$_2$ on grain surfaces under the physical conditions in 
the layers responsible for H$_2$ emission. 
We use steady state PDR models in 
order to examine the sensitivity of different H$_2$ line ratios to the H$_2$ 
formation rate $R_{f}$. We show that the ratio of the 0-0 S(3) to the 1-0 S(1) 
line increases with $R_{f}$ but that one requires independent estimates of the
radiation field incident upon the PDR and the density in order to infer $R_{f}$ 
from the H$_2$ line data. 
 We confirm the earlier result of \cite{habart2003a}
that the H$_2$ formation rate in regions of moderate excitation such as Oph W, 
S140 and IC 63 is a factor of 5 times larger than the standard rate 
inferred from UV observations of diffuse clouds. On the other hand, towards regions of higher 
radiation field such as the Orion Bar and NGC 2023, we derive H$_2$ formation
rates consistent with the standard value. 
We find also a correlation between the H$_2$ 1-0 S(1) line 
and PAH emission suggesting that $R_f$ scales with the PAH abundance.
With the aim of explaining these results, we consider some empirical models of the H$_2$ formation process.  Here we consider both formation on 
big ($a\sim$0.1 $\mu$m) and small ($a\sim$10 \AA) grains by either direct recombination from the gas phase 
or recombination of physisorbed H atoms with atoms in a chemisorbed site. 
We conclude that indirect chemisorption is most promising in PDRs.
Moreover small grains which dominate the total grain surface and spend most of 
their time at relatively low temperatures may be the most promising surface for 
forming H$_2$ in PDRs. 
}

\authorrunning{Habart et al.}
\titlerunning{H$_2$ formation rate in PDRs}

\maketitle

\section{Introduction}
The formation of molecular hydrogen is a key process affecting the thermal and density structure and the chemical evolution of the interstellar medium \cite[see, for example,][]{combes00}.
Although there is a consensus that H$_2$ forms on the surface of dust grains 
 \cite[]{gould63,Hollenbach71,jura75,duley84}, the mechanism is not yet understood.
This is partly due to our ignorance concerning interstellar grain composition, form, structure
 and physico-chemical state.
It is also caused by our lack of understanding of surface reactions in the 
interstellar context.

Numerous theoretical and experimental studies have thus been dedicated
to the study of the H$_2$ formation process \cite[]{sandford93,duley96,parneix98,pirronello97,pirronello99a,takahashi99,katz99,williams00,sidis2000,biham2001,joblin01,cazaux2002,cazaux2002a}. 
Another approach to this issue is to examine the H$_2$ formation rate in different regions of the ISM
 in order to see how it depends upon the local physical parameters.
One can for example study the correlation of H$_2$-related quantities (abundance, rotational excitation, vibrational excitation) with the local dust properties. 
 ISO observations of H$_2$ lines and small carbonaceous grains emission in several PDRs, and new H$_2$ UV 
absorption observations by FUSE of diffuse clouds, 
open new perspectives on our understanding of the H$_2$ formation process.
In particular, the confrontation between observations and theoretical 
predictions provides strong constraints upon the H$_2$ formation rate \cite[]{habart2003a,gry2002,tumlinson2002}.

In this paper, we consider PDRs and
investigate the H$_2$ formation process using observations
 of H$_2$ emission obtained by ISO and ground-based
telescopes. PDRs, where stellar radiation plays a dominant
role in determining the chemical and thermal state of the gas \cite[for a recent review see][]{hollenbach99}, are privileged objects 
for the study of physical and chemical processes of the interstellar medium. 
In Sect. \ref{objects}, we review the data available
for five nearby PDRs and summarize their physical conditions. In Sect. \ref{observations}, we estimate 
for each PDR of our sample the H$_2$ formation rate using the H$_2$ line intensity ratios as a diagnostic.
Also, in order to probe the influence of PAHs on the H$_2$ formation,
we compare in Sect. \ref{SGs_diagnostic} the intensities of emission measured in H$_2$ fluorescent lines with those in the aromatic bands.
Then, with the aim of explaining these results, we consider 
in Sect. \ref{mechanisms} empirical models of the H$_{2}$ formation
mechansim by recombination of an H atom in a bound site with
 a second H atom which is either in the gas phase or on a neighbouring
 physisorbed site.
In Sect. \ref{comparison}, we compare the observational constraints on the H$_2$ formation rate with different model predictions. Our conclusions are summarized in Sect. \ref{conclusion}. 

\section{Sample of PDRs observed by ISO}
\label{objects}

{\small
\begin{table*}[htbp] \caption{Sample of PDRs observed by ISO}
\label{pdrs}

\begin{tabular}{llllll}
\hline
                       & Oph W  & S140   & IC 63 & NGC 2023 & Orion Bar \\
\hline
\hline
$\chi$$^a$                & 250  & 100 - 250  & 650  &500 - 3000  & 5000 - 2.4 10$^4$\\
Ref.                      & (1a) & (2a)-(2b) & (3a) & (4a)-(4b) & (5a)-(5a,b)   \\
&  & &  & & \\
$n_H$$^b$ (cm$^{-3}$)     & 10$^4$ & 10$^4$ - 5 10$^4$ & 5 $10^4$ - $10^5$ & 10$^4$ - 10$^5$ & 5 10$^4$ - 3 10$^5$\\
Ref.                      & (1a) & (2b)-(2a) & (3a) & (4c)-(4a,b,d) & (5b,c,e)-(5e,f,g)\\
&  & &  & & \\
$A_V$$^c$ (mag)&  10  & $\gtrsim$10 & 7 & $\gtrsim$10 & $\gtrsim$10  \\
Ref.                           &  (1b)    &(2c)  & (3a)  & (4a)  & (5h)\\
\hline
\end{tabular}

$^a$ Incident FUV radiation field expressed in units $\chi$ of the \cite{draine78} average interstellar radiation field. \\
$^b$ Proton gas density $n_H\equiv n_{H^0}+2~n_{H_2}$. \\
$^c$ Visual extinction within the PDR inferred from sub-mm dust emission or CO observations. \\
References: (1a) \cite{habart2003a}; (1b) \cite{motte98}; (2a) \cite{wyrowski97a}; (2b) \cite{timmermann96}; (2c) \cite{minchin93}; (3a) \cite{jansen94,jansen95b}; (4a) \cite{wyrowski97a,wyrowski2000}; (4b) \cite{draine96,draine99a}; (4c) \cite{black87}; (4d) \cite{field98}; (5a) \cite{marconi98}; (5b) \cite{tielens85a}; (5c) \cite{tielens93}; (5d) \cite{tauber94}; (5e) \cite{wyrowski97}; (5f) \cite{herrmann97}; (5g) \cite{simon97}; (5h) \cite{hogerheijde95}.\\

\begin{tabular}{llllll}
& & & & & \\
\hline
                       & Oph W  & S140   & IC 63 & NGC 2023 & Orion Bar \\
\hline
\hline
H$_2$ 0-0 S(3)$^a$ 9.66 $\mu$m& 13.7[8]&16.8[30] &10[29]  & 16.5[13.5] &59.7[11]\\          
(SWS, 20'')& & & & & \\
$T_{rot}^b$ (K)       &    330$\pm$15       &500$\pm$40         &620$\pm$45         & 330$\pm$15 &390$\pm$20\\    
$N_{H_2}^c$ (10$^{21}$ cm$^{-2}$)& 0.6 & 0.2 & 5 & 0.7 & 1 \\ 
Ref.                  & (1)       & (2)    &(3a)     &(4a)        &(5a)\\    
& & & & & \\
H$_2$ 1-0 S(1)$^{a}$ 2.12 $\mu$m & 3.1[16]  & -  &1.84[13]  & 7[20]   &13[10]\\    
Beam                  &    1''    &       & 74''             &1''            &1.5''\\    
1-0 S(1)/2-1 S(1)             & -        &  - &     2.2[23]     & 2.8[18]       &2.3[7] \\    
Ref.                  & (1)       &     &(3b)     &(4b,c)        &(5b,c)\\    
& & & & & \\
PAH (2-15$\mu$m)/10$^3$ & 3.6 & - & 0.8  & 11 & 62\\    
(ISOCAM, 6'')  & & & & & \\ 
Ref.                  & (1)        &    & (3c)     &(4d)        &(5d) \\
\hline
\end{tabular}

$^a$ Intensities (in 10$^{-5}$ erg s$^{-1}$ cm$^{-2}$ sr$^{-1}$) with relative uncertainty in \% (in between brackets). H$_2$ line intensities
have not been corrected for dust attenuation. The 1-0 S(1) line intensity has been smoothed to the $\sim$20'' beam of SWS or multiplied by a beam factor.\\
$^b$ Excitation temperature of H$_2$ pure rotational levels with $J\le 7$. \\
$^c$ H$_2$ column density inferred from the intensity of H$_2$ rotational lines and assuming that the population distribution of low H$_2$ rotational levels is essentially in LTE. \\
References: (1) Data corresponding to the peak of the H$_2$ emission in \cite{habart2003a}; 
(2) \cite{timmermann96}; 
(3a) \cite{thi99} (note that the observations cannot be fitted by a 
single excitation temperature but two components at $\sim$100 K and $\sim$620 K are needed); (3b) \cite{luhman97a}; (3c) data from Cesarsky, priv. com.; 
(4a) \cite{moutou99}; (4b) \cite{field98}; (4c) \cite{burton98}; (4d) \cite{abergel2002};
(5a) data from Bertoldi, priv. com.; (5b) \cite{vanderwerf96}; (5c) \cite{walmsley2000}; (5d) \cite{cesarsky2000}.

\end{table*}

\normalsize

Recently, the Short Wavelength Spectrometer \cite[SWS,][]{Kessler96} on board ISO has observed a series of
pure H$_2$ rotational lines towards a variety of nearby PDRs.
In the ISO data base, we have selected five nearby PDRs which sample well the range of excitation conditions covered by the SWS observations. At the low excitation end, we have the Oph W and the S140 PDRs. 
Then, with radiation field higher by a factor of 3-10, 
we have the PDR IC 63 
and the reflection nebula NGC 2023 (where we focus on the filament
 at 60'' south of the central star). 
Finally, at the high excitation end, we have the Orion Bar at the position of the peak of the fluorescent H$_2$ emission. These PDRs close to the Sun ($d\sim$100-500 pc) are ideal targets to discuss the formation of H$_2$ in 
hot regions of the ISM.

In Table \ref{pdrs}, we summarize the physical conditions prevailing
 in each region as determined from the literature. 
The thermal and chemical structure of the PDR depends on two parameters,
 namely, the intensity of the incident 
far-ultraviolet (FUV, 6$< h\nu <$13.6 eV) radiation field and the
gas density. We adopt the radiation field of \cite{draine78}
and we characterize its intensity with a scaling factor $\chi$ ($\chi$=1
corresponds to the standard FUV interstellar radiation field of $2.6~10^{-3}$ erg s$^{-1}$ cm$^{-2}$).
This factor 
is determined from the expected FUV luminosity of the exciting star and assuming that the 
distance of the exciting star to the PDR is equal to the distance
projected onto the sky.
This in principle is an upper limit and hence we also report in Table \ref{pdrs} estimations of $\chi$ based on observations of the fine structure lines of C$^+$ or O$^0$. In general, we note that $\chi$ is uncertain by a factor of about 2 to 5.

The proton gas density $n_H$ has been derived from a variety of observational constraints:
(1) for Oph W, $n_H\sim 10^4$ cm$^{-3}$ is estimated from the brightness profile of the aromatic dust emission \cite[]{habart2003a}, while observations in the C$^0$ and CO lines suggest $n_H\sim 6~10^4$ cm$^{-3}$
 for the inner part of the PDR \cite[]{habart2001};
(2) for S140, using [C$^+$] 158 $\mu$m and C$^0$ radio recombination line,
\cite{wyrowski97a} derive a density about $n_H\sim 5~10^4$ cm$^{-3}$ comparable to the gas density expected for pressure balance with the HII region $\sim 10^4$ cm$^{-3}$ \cite[]{timmermann96};
(3) measurements of CO, HCO$^+$, HCN or CS line ratios from the PDR IC 63 suggest $n_H\sim 5~10^4-10^5$ cm$^{-3}$ \cite[]{jansen94,jansen95b}; 
(4) for the bright southern emission bar of NGC 2023, the inferred densities are in the range $\sim 10^4-10^5$ cm$^{-3}$ using H$_2$ fluorescent line emission 
\cite[]{black87,draine96,field98,draine99a} and $\sim 10^5$ cm$^{-3}$
from [C$^+$] 158 $\mu$m and C radio recombination line intensity \cite[]{wyrowski97a,wyrowski2000};
(5) finally, for the Orion Bar the gas density has been estimated 
to be about $\sim 5~10^4$ cm$^{-3}$ 
from the observed stratification of different tracers of PDRs \cite[]{tielens93}
and $\gtrsim 10^5$ cm$^{-3}$ from observations of fine-structure 
line emission \cite[]{herrmann97}, as well as C radio recombination lines \cite[]{wyrowski97} and CN, CS observations \cite[]{simon97}.
The gas density inferred from various atomic/molecular species shows a relatively large dispersion (typically from 10$^4$ to $10^5$ cm$^{-3}$). 
This dispersion could result from systematic density gradients from the H$_2$
emitting layer to the molecular cold layer \cite[see, for example,][]{walmsley2000,habart2003a}.
What is needed for our study is the density in the H$_2$ emitting region while mainly
 the density tracers reflect the density of the cold gas in the cloud.

We report also in Table \ref{pdrs} the observed intensities for the H$_2$ 0-0 S(3) and 1-0 S(1) lines
 obtained respectively with ISO-SWS and from ground based observations. High spatial resolution observations ($\sim$1'') of the 1-0 S(1) line emission have been smoothed to the $\sim$20'' beam of SWS. 
For measurements made with beams larger than the SWS observations, we have 
scaled the flux according to the beam ratio. For each region, we also give the excitation temperature 
of the H$_2$ pure rotational levels with $J\le 7$, the 
H$_2$ column density inferred from H$_2$ rotational lines intensity, and the 1-0 S(1)/2-1 S(1) line ratio. 
In the case of NGC 2023 and the Orion Bar, the 1-0 S(1)/2-1 S(1) line ratio
has been taken respectively at 18''S 11''W \cite[]{burton98} and 50''N 30''E \cite[]{walmsley2000}
 of the SWS pointings. Finally, based on ISOCAM observations\footnote{From the brightness in the LW2 filter (5-8.5 $\mu$m), which is dominated by the aromatic dust emission \cite[]{boulanger98}, we can estimate the aromatic dust emission using the following relationship : $I_{PAH}(2-15 \mu m) \simeq 2 \times \nu I_{\nu}(5-8.5 \mu m)$ based on ISOCAM-CVF spectrum (corrected from the dust continuum emission) taken in PDRs.}, we give the aromatic dust emission 
smoothed to the SWS beam. In the following,
aromatic dust particles will be hereafter referred to as PAHs (Polycyclic Aromatic Hydrocarbons).
 This is a generic term which encompasses large aromatic molecules and tiny carbonaceous dust grains containing up to a few 1000 atoms and with radii of a few \AA\ to a few tens of \AA.

\section{Estimates of the H$_2$ formation rate}
\label{observations}

In this section, we show how the H$_2$ formation rate can be derived from the analysis
of the H$_2$ emission line using PDR models.

For several PDRs observed by ISO, the H$_2$ line intensities
and the gas temperature as probed by the populations of the
low rotational levels of H$_2$ were found to be higher than 
predicted by current models \cite[]{bertoldi97,draine99,thi99,habart2003a,li2002}.
The cause of this discrepancy is that, in the models, the gas is not hot enough or alternatively that the
column density of H$_2$ is too low in the zones where the gas is warm.
One explanation of this discrepancy is that the H$_2$ formation rate is larger
 at high gas temperatures, moving the H$^0$/H$_2$ transition zone closer
to the edge of the PDR.  
\cite{habart2003a} have shown that the observed H$_2$ excitation from the 
 moderately excited Oph W PDR can be accounted for by increasing the H$_2$ formation rate by a factor about 5 \cite[compared to the standard H$_2$ formation rate derived in the diffuse ISM by][]{jura75} to $\sim 2~~10^{-16}$ cm$^{3}$ s$^{-1}$ at $T_{gas}\simeq$330 K. In this study, we extend our study of the H$_2$ formation 
in PDRs (described in Sect. \ref{objects}) spanning a wide range of excitation conditions.

\subsection{PDR model}
\label{model}

In order to analyse the H$_2$ emission observations from the PDRs,
we use an updated version of the stationary PDR model described in \cite{lebourlot93}.
In this model, a PDR is represented by a
semi-infinite plane-parallel slab with an incident radiation field. The input parameters are {\rm (i)} the incident FUV field $\chi$, and {\rm (ii)}
 the proton gas density $n_H$. 
We will consider constant density models with $n_H=10^4-10^5$ cm$^{-3}$
and $\chi=10^2-10^5$ in the range of values prevailing in our PDR sample 
(see Table \ref{pdrs}).
With the inputs $\chi$ and $n_H$, the model solves the chemical and thermal balance starting from the slab edge at each $A_v$-step in the cloud.
The H$_2$ abundance results from a balance between the formation of H$_2$ on dust grains and the photodissociation of H$_2$ by FUV flux, which is attenuated by dust extinction and self-shielding in the
H$_2$ lines. At equilibrium, the density of atomic ($n_{H^0}$) and molecular hydrogen ($n_{H_2}$) are given by
\begin{equation}
R_f~n_H~n_{H^0} = R_{d}(0) \times \chi e^{-\tau _d} \times f_{s}(N(H_2)) \times n_{H_2}
\label{Eq_Rf}
\end{equation}
with $R_f$ (cm$^3$ s$^{-1}$) the H$_2$ formation rate, $R_{d}(0)\sim 5\times 10^{-11}$ s$^{-1}$ 
the unshielded photodissociation rate per H$_2$ for $\chi$=1, 
$N(H_2)$ the colum density of H$_2$ and $e^{- \tau _{d}}$ and $f_{s}(N(H_2))$ respectively the dust extinction and the H$_2$ self-shielding factors.
In this study, we adopt $R_f$ constant throughout the PDRs.
The standard value inferred from observations of H$_2$
UV absorption lines in interstellar diffuse clouds
 is $R_f^0\sim 3~10^{-17}$ cm$^{3}$ s$^{-1}$
\cite[]{jura75}.
We assume that the energy released by the nascent molecule (4.5 eV) is 
equally distributed between the kinetic energy of H$_2$ ($E_k$), the internal
 energy of H$_2$ ($E_{int}$) and the internal energy of the grain ($E_{grain}$). 
Moreover, we assume that the internal energy of the nascent H$_2$ is distributed in Boltzmann distribution through the energy levels.
 The energy used for the heating per H$_2$ formation 
is $E_k + E_{int}\times f$ with $E_k\sim$1.5 eV and $E_{int} \times f~\sim$0.7 eV where $f$ is the fraction of the internal
energy of the nascent H$_2$ contributing to the heating.
For the collisional excitation and de-excitation of H$_2$, we adopt the
 H$^0$-H$_2$ inelastic rates 
 of \cite{martin95} extrapolated to low temperatures.
The heating rate due to the photoelectric effect on small dust grains is derived from the formalism of \cite{bakes94}.
For the physical conditions prevailing in our PDRs,
the heating is mainly due to the photoelectric effect.
Nevertheless, for high $R_f$ ($\sim 10 \times R_f^0$)
 the heating rate per H$_2$ formation becomes comparable
(factor of $\sim$2 lower) to the photoelectric
heating rate.

\subsection{H$_2$ line intensity ratios as diagnostic of the H$_2$ formation rate}
\label{Rh2_diagnostic}

The ($v$, $J$) excited states of H$_2$ can be populated by inelastic collisions with gas phase species, UV pumping and by the formation process.
The population distribution of H$_2$ levels is a function of the gas density, the gas temperature and the UV flux.  Hence, the H$_2$ line intensity ratios - which depend on the physical conditions in the 
photodissociation front where atomic hydrogen becomes molecular -  should probe the H$_2$ formation rate which controls the location of the H$^0$/H$_2$ transition zone in PDRs.
In the following, we study the dependence of several H$_2$ line ratios as 
a function of $R_f$ (predicted by model calculations).

In Fig. \ref{fig_h2_mod}, we show the model predictions (face-on) for the intensity of the 1-0 S(1) line as well as for several
   commonly observed line intensity ratios as a function of
   the molecular hydrogen formation rate. 
First, we discuss the dependence of the 1-0 S(1) line 
intensity on $R_f$.
At equilibrium (Eq. (\ref{Eq_Rf})),
the intensity of the 1-0 S(1) line $I^f_{H_2}$ is proportional to
$R_f n_H N(H^0)$ where $N(H^0)$ is the column density of atomic H atoms.
For high values of $\chi/n_H$ ($\gtrsim 0.01$), the 
H$^0$/H$_2$ transition is driven by the dust opacity
and $N(H^0)$ is a constant equal to a few $10^{21}$ cm$^{-2}$. 
We thus have $I^f_{H_2}\propto R_f n_H$ (see Fig. \ref{fig_h2_mod}(a)). 
Conversely, when $\chi/n_H\lesssim 0.01$, molecular hydrogen self shields sufficiently that the gas goes molecular before 
the dust gets optically thick.
In this case, using the approximation $f_{s}(N(H_2))=(N(H_2)/N_0)^{-3/4}$ with $N_0=10^{14}$ cm$^{-2}$ \cite[]{draine96} for $N(H_2) \lesssim 10^{21}$ cm$^{-2}$,
we find $R_f n_H N(H^0) = 4 R_{d}(0) \times \chi \times N(H_2)^{1/4} N_0^{3/4}$ \cite[]{hollenbach99}.
At the H$^0$/H$_2$ transition, we have $N(H_2)=N(H)/2$ with $N(H)$
the proton column density from the edge of the PDR to the H$^0$/H$_2$
transition zone. Finally, $I^f_{H_2}\propto \chi \times N(H)^{1/4}$ (see Fig. \ref{fig_h2_mod}(a)).

\begin{figure*}[htbp]
\leavevmode
\centerline{ \psfig{file=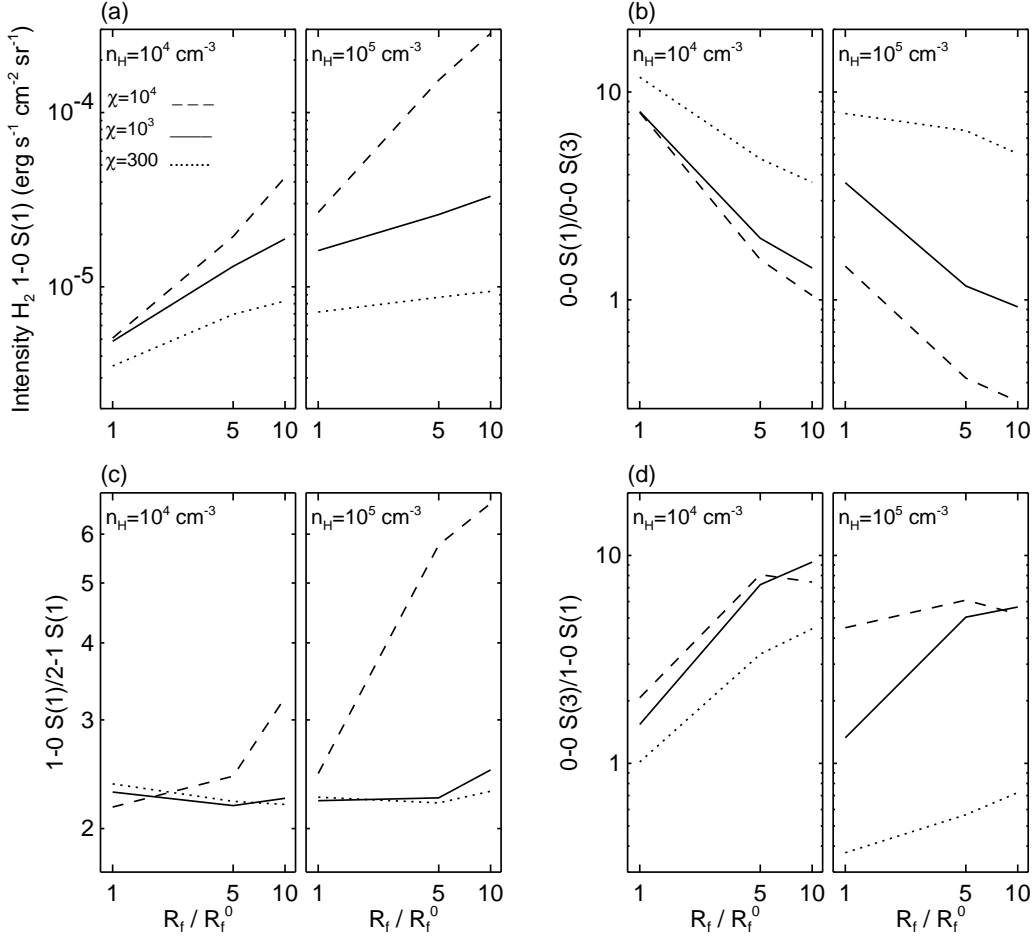,width=14cm,angle=0} }
\caption{\em 
Panel {\bf(a)} : Intensity of the H$_2$ 1-0 S(1) line emission predicted by models (face-on) for two gas densities
and different incident radiation fields - $\chi$=300 (dotted lines), $\chi$=10$^3$ (solid lines) and $\chi$=10$^4$ (dashed lines) - as a function of the value of H$_2$ formation rate
divided by $R_f^0$ corresponding to the formation rate based upon Copernicus data \cite[]{jura75}. Panels {\bf(b)}, {\bf(c)} and {\bf(d)}: several H$_2$ line intensity ratios for the
same models.}
\label{fig_h2_mod}
\end{figure*}

  The low-J rotational v=0-0 H$_2$ lines have a somewhat different dependence
  upon $R_{f}$ because at densities above $10^4$ cm$^{-3}$ of interest to
  us, their population distribution is essentially in LTE and  the
  line intensities depend mainly on the temperature in the
   photodissociation front where hydrogen becomes molecular.
   In the range of $\chi$ and $n_{H}$ studied here, the gas temperature
   profiles as a function of depth in the PDR are fairly  insensitive
   to the fraction of molecular hydrogen (heating is mainly due to photoelectric
   emission from dust grains and cooling to fine structure line emission
   of O$^0$ and C$^+$) but higher H$_2$ formation rates cause the photodissociation
   front to shift  closer to the surface where the temperature is higher.
   We  show in Figs. \ref{fig_h2_mod}(b) and \ref{fig_h2_mod}(d) 
the dependence of the 0-0 S(1)/0-0 S(3)
   and 0-0 S(3)/1-0 S(1) line ratios as functions of $R_{f}$.
   The ratio of the two 0-0 lines is essentially a measure of temperature
   and thus decreases as $R_{f}$ increases as a consequence of the 
temperature increase in the dissociation front.  This effect also causes a
   sharper increase in 0-0 S(3) than in 1-0 S(1) and thus the
   ratio 0-0 S(3)/1-0 S(1) increases with increasing $R_{f}$.

   We also show in Fig. \ref{fig_h2_mod}(c) the dependence of the 2-1 S(1)/1-0 S(1)
   line predicted by the models as a function of $R_{f}/R_f^0$.  This
   commonly observed line ratio is known in PDRs to vary between
   values of $\sim$2 typical of pure fluorescence to values of $\sim$3-5
   at high densities (above 10$^5$ cm$^{-3}$)
   when collisional deexcitation of higher vibrational
   levels becomes competitive with radiative decay
   \cite[]{black87,draine96}. One sees from
   Fig. \ref{fig_h2_mod}(c) that the model predicted line ratio 
 for $\chi \le 10^3$ and $n_H \le 10^5$ cm$^{-3}$ is independent of $R_{f}$ (as well as of $\chi$ and $n_H$) and equal to $\sim$2. 
 This is due to the fact that the densities and radiation fields of interest to us
    (and consequently assumed in the models) are in the range for which
the population of excited vibrational levels of H$_2$ are controlled by
purely radiative cascade process. 

In summary, we find that the 0-0 S(3)/1-0 S(1) line intensity ratio
 depends considerably on the H$_2$ formation rate via the variation of both the gas temperature and the UV flux in the H$^0$/H$_2$ transition zone. 
However, the changes in this ratio can also be due to changes in the radiation
field and density. Thus, by comparing this H$_2$ line intensity ratio
predicted by the model (essentially independent of the geometry and of the total
column density) with observations from PDRs where
$\chi$ and $n_H$ have been determined from other observations, we can expect 
to probe the H$_2$ formation rate in PDRs.

In Fig. \ref{fig_h2_obs}, we show the H$_2$ 0-0 S(3)/1-0 S(1)
line intensity ratio as predicted by the model for an edge-on geometry
(as most of our PDRs are,
see references in Table \ref{pdrs}) in the H$_2$ emission region
as a function of $\chi$ and for different H$_2$ formation rates.
We compare these predicted line intensity ratios with observational data for our PDRs sample described in Sect. \ref{objects}.
The data has been corrected for dust attenuation applying 
a correction factor $\tau/(1-e^{-\tau})$. 
Here, we use the extinction curve of \cite{draine89} and 
the visual extinction within the PDR reported in Table \ref{pdrs}
 derived from sub-mm dust emission or CO observations.
This assumes that our PDRs are exactly planar and edge-on  
and that densities are similar in the H$_2$ emission layer and
in the cold molecular layer which is questionable because
 the column density inferred from the molecular hydrogen 
rotational transitions is generally an order of magnitude lower than
the value derived from sub-mm dust emission or CO observations. 
The explanation of this difference is not clear. It could
occur in an edge-on geometry due to beam dilution effects (i.e., 
H$_2$ emission region not resolved by ISO-SWS) or because of
 density gradients. For reasons of simplicity,
we have adopted in Fig. \ref{fig_h2_obs} the estimate inferred 
from sub-mm dust emission or CO observations.
The error involved here is small since using the column density
derived from the molecular hydrogen rotational transitions,
 the H$_2$ 0-0 S(3)/1-0 S(1) line intensity ratio would
diminish by $\sim$30\% which is comparable to the error bars.

\begin{figure*}[htbp]
\leavevmode
\centerline{ \psfig{file=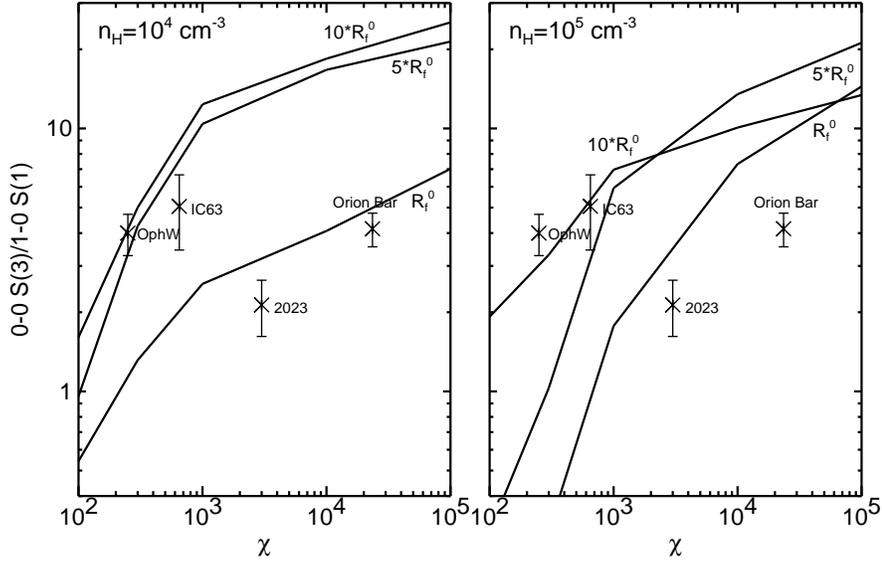,width=13cm,angle=0} }
\caption{\em H$_2$ 0-0 S(3)/1-0 S(1) line intensity ratio, as predicted by models (edge-on)
and as observed for our PDR sample (crosses with error
bars), as a function of $\chi$ the FUV incident radiation field. Predicted
line intensity ratios (taken in the H$_2$ emitting zone) are presented for two gas densities
and shown for three different H$_2$ formation 
rates ($R_f^0$, $5 \times R_f^0$, $10 \times R_f^0$).
Observed line intensity ratios have been corrected for dust attenuation
(see text). For each PDR of our sample, $\chi$ has been derived from the expected FUV luminosity of the exciting star and assuming that the 
distance of the exciting star to the PDR is equal to the distance
projected onto the sky (see Sect. \ref{objects}). For S140, where we have no measurement of the intensity for the 1-0 S(1) line,
we use observations of the 1-0 Q(3) line to determine the H$_2$ line intensity ratio (see text).}
\label{fig_h2_obs}
\end{figure*}

\nopagebreak

\begin{figure*}[htbp]
\leavevmode
\centerline{ \psfig{file=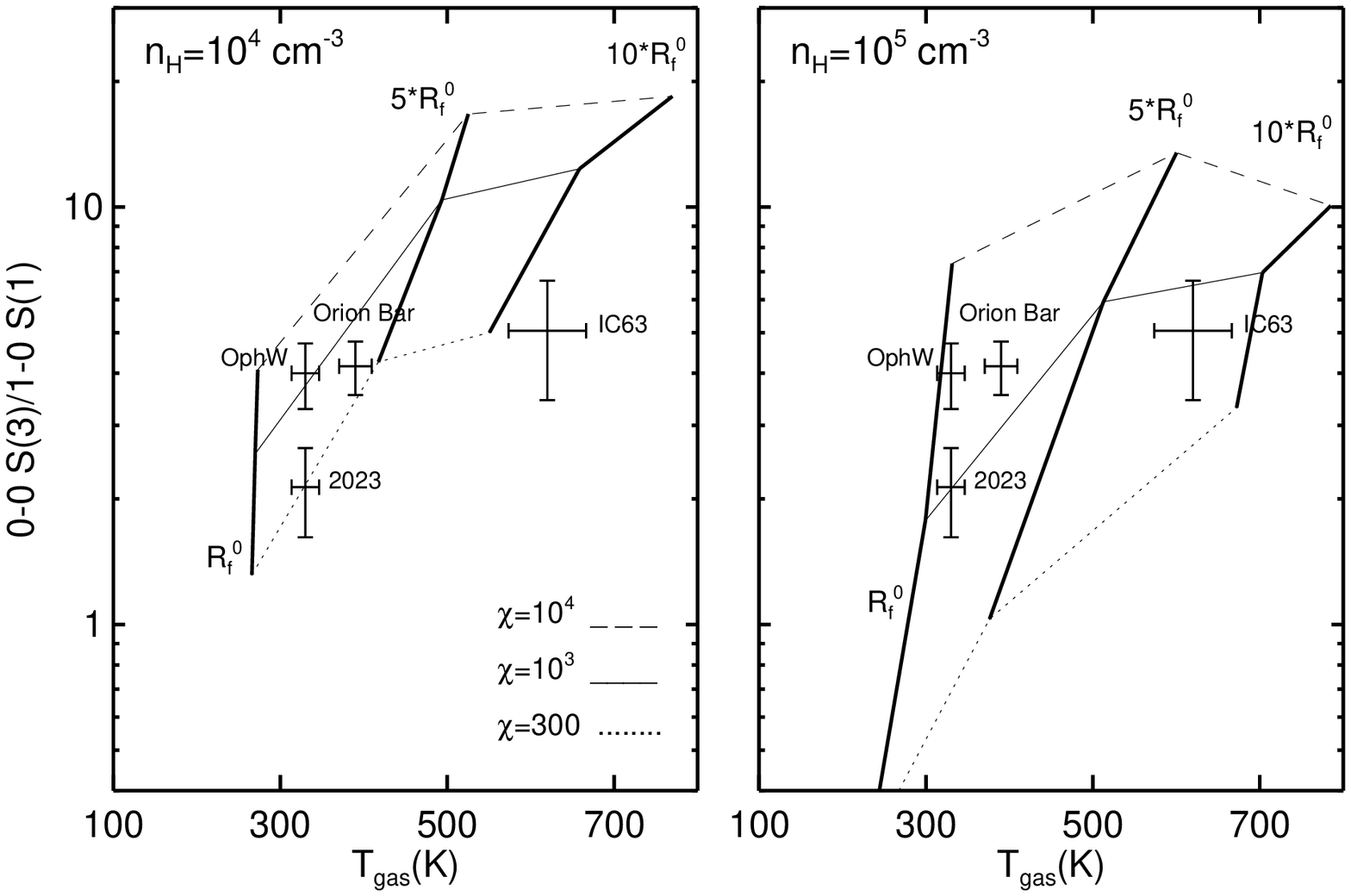,width=13cm,angle=0} }
\caption{\em H$_2$ 0-0 S(3)/1-0 S(1) 
line intensity ratio as predicted by the same models shown
 in Fig. \ref{fig_h2_obs} and as observed for our PDR
sample (crosses with error bars) as a function of $T_{gas}$ the gas temperature
 in the H$_2$ emitting region. For the observations, we use the
excitation temperature of the H$_2$ pure rotational levels (see text).
For IC 63, the observed H$_2$ pure rotational
lines intensity cannot be fitted by a 
single excitation temperature but two components at $\sim$100 K 
and $\sim$600 K are needed.}
\label{fig_h2_obs_Tg}
\end{figure*}

In the following, we discuss for each PDR of our sample the value of $R_f$ required to account
for the observed H$_2$ excitation. 
For Oph W ($\chi \sim$250, $n_H\sim$10$^4$ cm$^{-3}$) and IC 63 ($\chi \sim$650, $n_H\sim 5~10^4 - 10^5$ cm$^{-3}$), we find that models with a high H$_2$ formation rate (i.e., $R_f\gtrsim 5\times R_f^0$) roughly reproduce the 0-0 S(3)/1-0 S(1) line ratio observed.
For the standard H$_2$ formation rate, the 0-0 S(3)/1-0 S(1) line ratio
is underestimated by a factor about 4.
For the more highly excited PDRs, i.e., NGC 2023 ($\chi \sim 500-3000$, $n_H\sim 10^4-10^5$ cm$^{-3}$) and the Orion Bar ($\chi \sim 0.5-2.5~10^4$, $n_H\sim 5~10^4-3~10^5$ cm$^{-3}$), 
we find on the contrary that models with the standard H$_2$ formation rate roughly match the data. 
In the case of S140, where we have no measurement of the intensity for the 1-0 S(1) line,
we use ISO observations of the 1-0 Q(3) line from \cite{draine99} to determine the H$_2$ rotational to rovibrational line intensity ratio. For the physical conditions of interest to us, the models predict that the
 1-0 S(1)/1-0 Q(3) line intensity ratio is about a factor of 1.4.
Applying this factor, we find that the 0-0 S(3)/1-0 S(1) line intensity ratio (corrected
for dust attenuation) is about 3. Then, one sees from Fig. \ref{fig_h2_obs}
that for S140 ($\chi \sim$100-250, $n_H\sim 10^4-5~10^4$ cm$^{-3}$) 
models with a high H$_2$ formation rate are - as for the moderately excited PDR Oph W and IC 63 -
 required to explain
the observed H$_2$ line intensity ratio.
 In Table \ref{rate_formation}, we give for each region the values of the H$_2$ formation rate 
derived using Fig. \ref{fig_h2_obs} and assuming that the gas density in the
H$_2$ emission region is about $10^4$ cm$^{-3}$ for all sources.
Considering that the gas density
is about $10^5$ cm$^{-3}$, $R_f$ required to explain the data would be (by a factor of $\sim$2)
 higher and lower for respectively the moderately and the highly excited PDRs.
We emphasize that our determination of $R_f$ from the H$_2$ line intensity ratios
depends on the treatment in the model of the thermal balances and the chemistry.
In particular, the calculation of the grain photoelectric heating
which mainly determines $T_{gas}$ in the PDR modelling (see Sect. \ref{model}) could 
affect the excitation of the lowest rotational levels ($J\le$5) of H$_2$.
For example, \cite{habart2003a} have shown 
that, throughout the Oph W PDR, increasing 
the photoelectric heating rate by $\sim$50\%,
which leads to higher $T_{gas}$ in the H$^0$/H$_2$ transition,
the 0-0 S(3) line intensity is enhanced (by a factor $\sim$2) and 
$R_f$ required to explain the data would be reduced by a factor $\sim 1.5$ (see Fig. \ref{fig_h2_obs}).
However, even if we cannot precisely determine the uncertainties on our estimates 
of $R_f$, the values of $R_f$ derived here should be at most uncertain by a factor of about 2.

One check of our estimates of $R_f$ is to see if the
predicted $T_{gas}$ values in the H$_2$ emitting region
are consistent with the observed rotational temperatures reported in Table
\ref{pdrs}. In Fig. \ref{fig_h2_obs_Tg}, we show the 0-0 S(3)/1-0 S(1) 
line intensity ratio predicted by the same models shown
 in Fig. \ref{fig_h2_obs} as a function of the gas temperature
 in the H$_2$ emitting region and compared with observational data. 
 For observations, we use the
excitation temperature of H$_2$ pure rotational levels (with $J\le 7$)
 which in principle is an upper limit as UV pumping
could contribute to the excitation of H$_2$ even for
 low energy levels.
In the case of IC 63 where the measured pure H$_2$ rotational
lines show two excitation temperatures \cite[at $\sim$100 K and
$\sim$600 K,][]{thi99} there is direct evidence for such contamination.
Using Fig. \ref{fig_h2_obs_Tg}, we favour (as previously)
 for the moderately excited PDRs models with high H$_2$ 
formation rate which can in fact explain the observed
 rotation temperatures.
Models with a standard H$_2$ formation rate predict in fact 
a gas temperature ($\le$300 K) lower than observed ($\sim$300-600 K).
For the Orion Bar and NGC 2023, we find on the contrary that models 
with standard H$_2$ formation rate predict temperatures consistent with 
the data ($\sim$300-400 K).
Finally, we have checked that these models reproduce the observed absolute intensities of the 0-0 S(3) and 1-0 S(1) H$_2$ lines.
Taking into account the inclinations of the PDRs of our sample, which are
seen edge-on except for IC 63, we find an agreement within $\lesssim$50\%.

It must be emphasized that these results are based on two fundamental assumptions. 
Firstly, we assume a uniform homogeneous gas density while 
PDRs may have a density gradient and could be clumpy. 
Nevertheless, because the modelling of the H$_2$ excitation mainly
depends upon the average gas density in the H$_2$ emission zone, density structure effects
are probably minor. 
In fact, the detailed study of \cite{habart2003a} of the H$_2$ excitation toward Oph W taking into account the gas density profile reaches the same estimation of
$R_f$ deduced here.

Secondly, we assumed a static, equilibrium PDR. In reality, the propagation of 
the ionization and photodissociation fronts will bring fresh H$_2$ into the zone 
emitting line radiation. Note that a non-equilibrium ortho-to-para
H$_2$ ratio has been observed in ISO-SWS observations 
towards Oph W \cite[]{habart2003a}, NGC 2023 \cite[]{moutou99} and
 NGC 7023 \cite[]{fuente99}. Nevertheless, the model of non-equilibrium
PDRs of \cite{stoerzer98} predicts that, for the physical conditions
prevailing in the Orion Bar and for an advection velocity of the
order of 1 km s$^{-1}$, the H$_2$ 0-0 S(3)/1-0 S(1) line intensity ratio
varies by a factor less than $\sim$2 relative to the steady state value. 

In the next section, we review other observational constraints on the H$_2$ formation rate based on H$_2$ UV absorption measurements. 

\subsection{Other observational constraints on the H$_2$ formation rate}
\label{rate_obs}

UV absorption measurements of H$_2$ have been often used to study H$_2$ formation, destruction, and excitation in the diffuse ISM. 
With the numerous new H$_2$ UV absorption line observations obtained recently by FUSE
\cite[]{snow2000,shull2000,rachford2001,rachford2002,tumlinson2002,sonnentrucker2002},
the H$_2$ formation in the diffuse ISM has been re-considered. 
In particular, \cite{gry2002} have determined the H$_2$ formation rates over three lines of
 sight in the Chamaeleon clouds. They find a rate
roughly constant and equal to about 4 10$^{-17}$ cm$^{3}$ s$^{-1}$ (with an uncertainty of about a factor of
2) for $n_H\sim 30-50$ cm$^{-3}$ and $T_{gas}\sim 60$ K, in agreement with the rate inferred by \cite{jura75}. Moreover, due to the high FUSE sensitivity, fainter stars with higher extinctions could be observed with far-UV instruments \cite[]{moos2000}, thus allowing the study of translucent clouds. \cite{rachford2002}, studying correlations of the H$_2$-related quantities with the column densities of other molecules and dust extinction properties for lines of sight with $A_v \gtrsim 1$, investigate the dependence of
 the H$_2$ formation rate with the composition and physical state of the gas and grains.
Furthermore, the FUSE observations in the lower metallicity environments
of the Small and Large Magellanic Clouds allow us to probe H$_2$ formation and destruction
in physical and chemical environments different from the Galaxy.
\cite{tumlinson2002} find that to reproduce the reduced molecular fraction
and enhanced rotational excitation in the SMC and LMC, a low H$_2$ formation
rate ($R_f\sim 3~10^{-18}$ cm$^3$ s$^{-1}$) and a high UV field relative to diffuse Galactic medium are required. 

Combining the FUSE results in the diffuse Chamaeleon clouds and the ISO observations discussed above,
we find that for a wide range of physical conditions - $1\lesssim \chi \lesssim 10^4$, 
100 cm$^{-3} \lesssim n_H \lesssim$ $10^5$ cm$^{-3}$, 50 K$\lesssim T_{gas} \lesssim$600 K,
 10 K$\lesssim T_{dust} \lesssim$100 K (see Tables \ref{pdrs} and \ref{rate_formation}) - H$_2$ forms efficiently ($R_f\sim$4 10$^{-17}$- 1.5 10$^{-16}$
cm$^3$ s$^{-1}$).
This result raises questions about our understanding of the
H$_2$ formation process. 
In Sect. \ref{mechanisms}, we re-examine models of H$_2$ formation with this in mind
and compare in Sect. \ref{comparison} the observational constraints on the H$_2$ formation rate with model predictions.

\section{Influence of aromatic dust on the H$_2$ formation process in PDRs}
\label{SGs_diagnostic}

In PDRs, small dust grains are intimately coupled to the evolution of the gas.
In fact, recent theoretical \cite[]{bakes94,weingartner01} and observational \cite[]{habart2001a} work
has shown that small grains (radius $\le$ 100 \AA) dominate the photoelectric heating. Furthermore, given that small grains make a dominant contribution to the total grain surface (see Sect. \ref{comparison}), it is plausible that they play a dominant role in H$_2$ formation. Therefore, if H$_2$ forms on small grains, we can expect that both the grain photoelectric heating rate and 
the H$_2$ formation rate will scale with their abundance. In particular, increasing their 
abundance would lead to higher gas temperature and would bring
the H$^0$/H$_2$ transition closer to the edge: this could significantly enhance the rotational and fluorescent emission of H$_2$. \cite{habart2003a} have shown that, assuming that $R_f$ scales with 
the PAH abundance, the enhancement in $R_f$ required to account for the observed H$_2$ emission from Oph W may result from an increased PAH abundance. Based on this, we investigate what we can predict concerning the influence of PAHs on the H$_2$ formation process.

To free oneself of the influence of the PAHs on the thermal balance, 
we study here the ratio between the H$_2$ 1-0 S(1) line (whose intensity does not depend
on $T_{gas}$, see Sect. \ref{Rh2_diagnostic}) and the PAH emission.
We first express analytically the H$_2$ fluorescent to PAH emission ratio 
$I^f_{H_2}/I_{PAH}$ as a function of the H$_2$ formation rate and the PAH abundance,
 as well as the physical conditions ($n_H$, $\chi$),
in order to examine qualitatively the dependence of this ratio with these parameters.
In Sect. \ref{Rh2_diagnostic}, we have seen that the H$_2$ 1-0 S(1) line
emission $I^f_{H_2}$
goes roughly as (i) $R_f n_H$ for high $\chi/n_H$ and as (ii) $\chi$ for low $\chi/n_H$.
The emission of aromatic dust scales with the intensity of the FUV radiation field \cite[]{puget85,sellgren85} and we have $I_{PAH} \propto \chi ~e^{- \tau _{d}} \times N_H~[C/H]_{PAH}$ with $[C/H]_{PAH}$ the abundance of carbon locked up in PAHs. As the column density over which 
PAHs emits is a few $10^{21}$ cm$^{-2}$ (where dust is optically
thin), $I_{PAH}$ goes as $\chi \times [C/H]_{PAH}$.
From these considerations, we find

\begin{equation}
\label{Eq:H2/PAH}
\frac{I^f_{H_2}}{I_{PAH}} \propto \left\{ \begin{array}{r@{\quad:\quad}l}
\frac{R_f}{[C/H]_{PAH}} \times \frac{n_H}{\chi} & \chi/n_H \gtrsim 0.01  \\ \frac{1}{[C/H]_{PAH}} & \chi/n_H \lesssim 0.01
\end{array} \right.
\end{equation}

Thus, by studying the observed H$_2$ fluorescent to PAH emission ratio
 from PDRs covering a wide range of $n_H/\chi$ ratio,
 we should be able to probe the H$_2$ formation on PAH: if H$_2$ forms on PAHs, we expect that the 
$I^f_{H_2}/I_{PAH}$ ratio scales with $n_H/\chi$. In other words, the $R_f/[C/H]_{PAH}$
 ratio is constant and does not vary from one region to another.

\begin{figure}[htbp]
\begin{center}
\leavevmode
\centerline{ \psfig{file=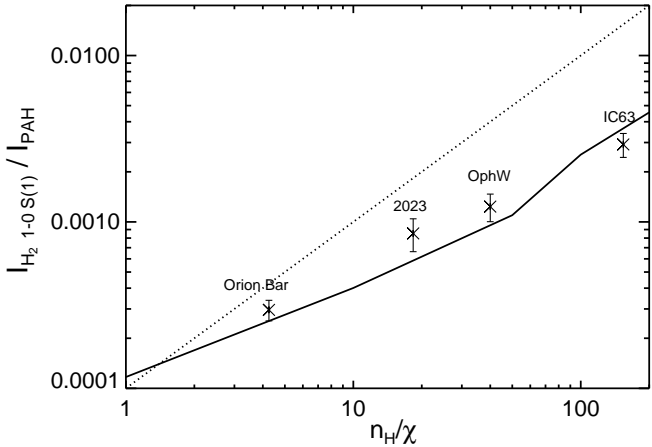,width=9cm,angle=0} }
\end{center}
\caption{\em  Ratio between the H$_2$ 1-0 S(1) line and PAH emission as predicted 
by edge-on model (solid line) and as observed (crosses
with error bars) as a function of $n_H/\chi$.
The dotted straight lines shows the linear dependence of the emission ratio with $n_H/\chi$ expected from Eq. (\ref{Eq:H2/PAH}) when $\chi/n_H \gtrsim 0.01$. For the PDRs of our sample, $n_H/\chi$ has been taken equal to the ratio
between the average value of the $n_H$ determinations given in Table \ref{pdrs}
and $\chi$ determined from the expected FUV luminosity of the exciting star and assuming that the 
distance of the exciting star to the PDR is equal to the distance
projected onto the sky (see Sect. \ref{objects}).}
\label{SGs_H2}
\end{figure}

We now compare the ratio between the H$_2$ 1-0 S(1) line and PAH emission
observed from the PDRs of our sample to model results.
For the models, we consider $\chi =10^{2}-10^{4}$ and $n_H=10^4-10^5$ cm$^{-3}$ which
 corresponds to the range of values prevailing in our PDR sample (see Table \ref{pdrs}). 
We adopt a constant $R_f/[C/H]_{PAH}$ ratio corresponding to
the values derived in the Oph W PDR, i.e., $R_f\simeq5\times R_f^0$ and $[C/H]_{PAH}\simeq$0.5 10$^{-4}$ \cite[]{habart2003a}. 
The power emitted by PAHs has been derived from the PAH absorption cross-section of \cite{verstraete92} and the size distribution described in Sect. \ref{comparison}. 
 In Fig. \ref{SGs_H2}, we show the H$_2$ 1-0 S(1)/PAH emission ratio predicted by the model in the H$_2$ emission zone as a function of the $n_H/\chi$ ratio.
As expected from the Eq. (\ref{Eq:H2/PAH}), for $\chi/n_H \gtrsim 0.01$ the H$_2$ 1-0 S(1)/PAH emission ratio increases proportionally with $n_H/\chi$. 

The observational ratios for our PDR sample are compared to these predictions. 
The observed points roughly fall on the model curve. 
From one PDR to another, the change of the emission ratio seems to result mainly from the physical conditions (i.e., $n_H/\chi$) variations.
We deduce from this that the $R_f/[C/H]_{PAH}$ ratio is roughly constant which suggests that formation of H$_2$ on PAHs should be important.
However, the correlation between $R_f$ and $[C/H]_{PAH}$ found
in our PDR sample is observed not to apply to the main PDR in the 30 Dor star forming region in the Large Magellanic 
Cloud where for $n_H/\chi\sim 1$ the H$_2$ 1-0 S(1) line to PAH 
emission ratio is measured to be $\sim$0.001 \cite[]{boulanger2003}. In future work,
this comparison should 
be extended so as to include all small grains, i.e., not only the band
carriers but also the very small grains emitting 
at longer wavelengths.

\section{H$_2$ formation mechanisms on grain surfaces}
\label{mechanisms}

In this section, with the aim of explaining our observational
constraints on the H$_2$ formation, we examine models
of H$_2$ formation mechanism.
Two general mechanisms for forming H$_{2}$ on grain
 surfaces have generally been proposed:

\begin{enumerate}
\item formation by physisorbed H atoms (Langmuir-Hinshelwood) whereby
two adsorbed and mobile physisorbed H atoms \cite[bound simply via van der Waals interaction
 with binding energies of the order of 0.05 eV or 500~K,][]{katz99} interact to recombine and desorb as H$_2$
and
\item formation by the interaction of an H 
atom from the gas phase with a chemisorbed H atom \cite[with binding energy roughly
 1 eV equivalent to $\sim$10,000~K,][]{fromherz93} forming desorbed H$_2$
(the Eley--Rideal mechanism).
\end{enumerate}

In the following discussion, we describe these processes mainly focusing 
on the version of the Langmuir--Hinshelwood mechanism whereby one of the
adsorbed H-atoms is originally in a chemisorbed site \cite[]{Hollenbach71,cazaux2002,cazaux2002a}. This 
approach will be called in the following the indirect chemisorption approach as opposed to the direct chemisorption or
Eley--Rideal mechanism. We make simple empirical estimates of rates for these processes
as a function of
the physical conditions and  grain characteristics. We consider small 
grains (SGs), dust particles with radii of a few \AA\ to 100 \AA\ (i.e., PAHs and Very Small Grains, VSGs) which fluctuate in temperature
in the radiation field,
and big grains (BGs), larger grains of radii $>$0.01 $\mu$m in thermal 
equilibrium with the radiation field \cite[]{desert90}.

\subsection{The Langmuir-Hinshelwood mechanism}
\label{for_phys}

In the formation from physisorbed atoms (pure Langmuir-Hinshelwood), 
an H atom is 
already physisorbed
on the grain surface. A second H atom from the gas sticks to the grain
and diffuses to finally recombine with the first H atom to form H$_2$. 
The rate depends upon
the competition between the mobility of the H 
atom on the grain surface and its thermal evaporation rate. 
In the laboratory, this process is observed to be efficient on grains with $7\lesssim T_{dust}\lesssim 20$ K 
\cite[]{pirronello97,pirronello99a}.

Amongst the dust grain populations described above, only BGs could
 carry a large fraction of physisorbed H atoms.
Indeed, after a UV photon is absorbed by SGs, the physisorbed H atoms should 
evaporate. Further, for the physical conditions typical of the PDRs,
the rate of thermal fluctuations $\tau _{abs}^{-1}$ \cite[$\sim 
1.6~10^{-9} \times N(C) \times \chi$ s$^{-1}$ with $N(C)$ the number of carbon 
atom in the grain,][]{verstraete2001}
is comparable to or larger than the accretion rate of H atoms $\tau 
_{acc}^{-1}$:
for a grain with a radius of 1.5 nm ($N(C) \simeq 1600$), we find $\tau _{abs}^{-1}/\tau 
_{acc}^{-1} \sim 1000 \times (100~K / T_{gas})^{0.5} \times \chi/n_H$.
Thus for $\chi/n_H$ larger than 0.001 (true in general for the
objects in Table \ref{pdrs}), UV radiation in PDRs will keep SG surfaces clean. 

Since this mechanism will only be efficient on BGs and in interstellar 
regions 
of low excitation (in PDRs we find typically $T_{BGs} \ge 30$K, see Table 
\ref{rate_formation}), we neglect it in what follows.

\subsection{Formation involving a chemisorbed H atom}
\label{for_chem}

We now consider the case where  the first
H atom is bound to the surface in a chemisorbed site. 
The second H atom from the gas phase reaches this site directly 
\cite[{\it direct} chemisorption or Eley-Rideal,][]{duley96,parneix98} or after 
diffusion on the grain surface \cite[{\it indirect} chemisorption,][]{Hollenbach71,cazaux2002,cazaux2002a}.
The incident H atom crosses  the activation barrier $E_a$ \cite[typically about $\sim$500-1500 K,][]{fromherz93,parneix98,sidis2000,cazaux2002a}
by either thermal hopping or  tunneling and recombines
to finally form H$_2$. This process
involving strongly bound H atoms will be efficient at higher dust surface 
temperatures than the pure Langmuir--Hinshelwood mechanism. 

We first consider the direct Eley--Rideal mechanism. 
 This
 involves reaction of the incoming 
H atom with an chemisorbed H on the grain surface without any requirement for adsorption 
and thermal accommodation. The formation probability $P_{H_2}$ can be written 
as: 
\begin{equation}
P_{H_2}=f \times \eta
\label{Eq_f_d_che_1}
\end{equation}
with $f$ the probability that the incident H atom reaches the chemisorbed H 
and recombines to form H$_2$ and $\eta$ the probability to eject the
molecular hydrogen formed. The incident H atom from the gas phase must
hit the grain with enough thermal energy to cross the activation barrier ($E_a$) by thermal hopping.
 $f$ is given by :
\begin{equation}
f=\frac{N_c}{N} \times \exp{\left(-\frac{E_a}{k T_{gas}}\right)} 
\label{Eq_f_d_che_2}
\end{equation}
with $N_c$ the total number of chemisorbed H atoms on the grain surface
and  $N$ the total number of physisorbed and chemisorbed sites.
The fraction of chemisorbed H atoms ($N_c/N$) is critical for this process. This fraction will be significant
when $T_{gas}$ is 
sufficiently high to carry an appreciable fraction of chemisorbed sites
occupied.
In PDRs (where $T_{gas}\ge$300 K) we expect $N_c/N$ 
to be relatively high while in cold interstellar clouds $N_c/N$ should be low.

An alternative to the above is the case where the incident H atom 
  initially sticks to a physisorbed site and then scans the surface until it
  ``finds'' the chemisorbed  H atom.  Here, the formation
probability $P_{H_2}$ can be written as: 
\begin{equation}
P_{H_2}=S \times f \times \eta
\label{Eq_f_ind_che_1}
\end{equation}
with $S$ the sticking probability. 

 The probability $f$ that the physisorbed
H atom diffuses over the grain surface (until it reaches a neighbouring site
to a chemisorbed H atom) and recombines before it is evaporated, can be given by
$f=\frac{\tau _{p}^{-1}}{\tau _{p}^{-1}+\tau _{ev}^{-1}}$
where $\tau _{p}^{-1}$ is the inverse time scale for diffusing and recombining and $\tau _{ev}^{-1}$ the evaporation rate of 
physisorbed H. In the appendix we estimate the time scales $\tau _{p}$ and $\tau _{ev}$. This leads to the expression:
\begin{equation}
f=\frac{1}{1+(\frac{N}{4N_c})^2 \times \exp{\left(\frac{E_p-E_d}{kT_{dust}}\right)}+(\frac{N}{4N_c})\times k_{rec}}
\label{Eq:f_ind_che}
\end{equation}
with  $E_d$ the desorption energy of a physisorbed H atom and $E_p$ the activation barrier energy for the diffusion of a physisorbed H (see Table \ref{parameter}). 
Considering that the physisorbed H atom crosses the activation barrier to recombine
with a neighbouring chemisorbed atom by either (i) thermal diffusion or by (ii) tunnelling, we find (see appendix)
that $k_{rec}=\exp{\left(\frac{E_a-E_d}{kT_{dust}}\right)}$ at high dust temperatures (above $\sim$40 K) when thermal diffusion
dominates whereas at low temperatures
(below $\sim$ 30 K) when tunneling is more important $k_{rec}=\exp{\left(E_d(\frac{1}{kT_{cr}}-\frac{1}{kT_{dust}})\right)}$ with $kT_{cr}=80~E_{d3}/(E_{a3}^{0.5}\times \Delta x(\AA))$ where 
$E_{a3}=E_a$/(1000 K), $E_{d3}=E_d$/(1000 K) and $\Delta x$ is the width of 
the barrier typically of the order of $\sim$1-3 \AA\ \cite[]{buch86,fromherz93,sidis2000}.

The activation barrier energies for physisorbed H atom diffusion ($E_p$) and
 for recombination with a neighbouring chemisorbed atom ($E_a$)
 are critical for this process.
For $E_a$ high ($\gtrsim$1000 K), the recombination term will dominate and $f$ will be high only where tunneling dominates (see Fig. \ref{h2_formation}).
For $E_a$ low ($\sim$600 K), the diffusion term will be critical and $f$ will
depend much on $E_p$. 
In this case, $f$ will
be high for $E_p\lesssim E_d$ (if $E_p$ is too large the atoms evaporate before
they find a neighbouring chemisorbed site) and for $E_p\sim 500$ K we find that the fraction of chemisorbed H ($N_c/N$) needs to be at least about 0.1 (see Sect. \ref{comparison}).

\begin{table} \caption{Parameters for carbonaceous dust surface.}
\label{parameter}
\begin{tabular}{lll}
\hline
Parameter & Value & Ref.\\
\hline
\hline
$E_d$(K) & 600 & (1)\\
$E_p$(K) & 500 & (1)\\
$E_a$(K) & 600-1000 & (2)-(3,4)\\
$\Delta x$(\AA) & $\sim$2 & (2)\\
\hline
\end{tabular}

$E_d$ is the desorption energy of a physisorbed H atom; $E_p$ is the activation barrier energy
for physisorbed H diffusion; $E_a$ and $\Delta x$ are respectively the energy and the width of 
the activation barrier for recombination with a neighbouring chemisorbed H.
Here, we assume that the barrier $E_a$ for recombination to 
form H$_2$ is the same between a physisorbed and a chemisorbed site as for direct recombination from the gas phase. \\
References. (1) \cite{katz99}; (2) \cite{cazaux2002a}; (3) \cite{fromherz93}; (4) \cite{parneix98} \\
\end{table}

We note that formation by chemisorption can happen on the surface of big or small
grains. In fact, the thermal fluctuations undergone  by SGs 
after absorbing a UV photon are not 
sufficient to evaporate 
chemisorbed H atoms (binding energy around 10,000 K). Moreover, 
 we estimate that the timescale for a physisorbed H to find a chemisorbed H is much less than the timescale of thermal fluctuations 
(see Sect. \ref{for_phys}).

Since SGs make a dominant 
contribution to the total
grain surface (see Sect. \ref{comparison}) and also because they may have numerous chemically bonded hydrogen 
atoms (because of their 
small size they are more disordered),
it is plausible that they play a dominant role in H$_2$ formation by 
chemisorption.
However, if small grains contribute efficiently to  H$_2$ formation, 
they must
be continuously rehydrogenated : this could be the case for larger PAHs and VSGs
for which the adsorbed H atoms should evaporate less after UV photon
 absorption because of the lower temperature fluctuations \cite[see, e.g.,][]{verstraete2001}.
 
In the next section, we compare the H$_2$ formation rates as predicted
by the different mechanisms discussed here and as derived from observations.

\section{Comparison between the H$_2$ formation rate as predicted and as derived from observations}
\label{comparison}

In this section, we see how the observational values of H$_2$ formation rate in PDRs
fit and constrain H$_2$ formation mechanisms on grain surfaces.

First, based on the empirical formation probabilities 
given above, we estimate the rate of  formation through chemisorption
on small and big carbonaceous grains. The formation rate $R_f$ (in cm$^{3}$ 
s$^{-1}$) can be written as:
\begin{equation}
R_f = \frac{1}{2} \sigma v_{H} \times P_{H_2}
\label{Eq:formation_h2}
\end{equation}
where $\sigma v_H$ is the collision frequency between H atoms (with mean 
velocity $v_H$) and grains and $P_{H_2}$ the probability of formation (see Eqs. 
(\ref{Eq_f_d_che_1}) and (\ref{Eq_f_ind_che_1})). The mean cross section for 
collisions between grains and H atoms, 
$\sigma$ (cm$^{2}$), is given by $<\frac{n_{grain}(a)}{n_H}\pi a^2>$ with
 $n_{grain}(a)$ the grain density with radius between $a$ and 
$a$+$da$.
We assume a MRN size distribution with $\alpha$=3.5 while the lower and upper 
limits
of the grain radius are respectively $a_{min}$=4 and $a_{max}$=100 \AA\ for SGs and $a_{min}$=0.01 and $a_{max}$=0.1 $\mu$m for 
BGs. Adopting uniform mass density $\rho$=2.25 g cm$^{-3}$
(typical value for graphite grains) and an grain/gas mass ratio $G$ about 0.001\footnote{The carbon locked up in SGs
has an abundance of $[C/H]\sim 10^{-4}$ inferred from comparison between observations of dust galactic emission and extinction with model calculations \cite[]{desert90,li2001a}.} for SGs and 0.01 for BGs,
  we estimate\footnote{We calculate $\sigma v_H$ with 
$v_H=(\frac{8kT_{gas}}{\pi m_H})^{1/2}$ and $\sigma=\int^{a_{max}}_{a_{min}} \pi a^2 a^{-\alpha} da \times \frac{V_{grain/gas}}{\int^{a_{max}}_{a_{min}} \frac{4}{3} \pi a^3 a^{-\alpha} da}$ with $V_{grain/gas}=1.4~m_H~G / \rho$ the volume of grain per hydrogen nuclei.} :
(i) for SGs, $\sigma v_H \sim 
6~10^{-16} (\frac{T_{gas}}{100~{\rm K}})^{0.5}~{\rm cm}^3~{\rm s}^{-1}$; (ii) 
for BGs, $\sigma v_H\sim 3~10^{-16} (\frac{T_{gas}}{100~{\rm 
K}})^{0.5}~{\rm cm}^3~{\rm s}^{-1}$. Under these assumptions,
we find that the collisions cross section is $\sim$2 higher for SGs than for BGs and that there is enough grain area with SGs to make $R_f$ larger than the standard H$_2$ formation rate. We assume the small dust abundance to be constant from one region to another although variations are expected.
However, this assumption should not be critical for our approach which attempts with simple empirical models to see what type of H$_2$ formation processes are to a first order relevant in PDRs.

To calculate $P_{H_2}$, we adopt the following assumptions.
Firstly, for the indirect chemisorption, the formation
efficiency will depend on the sticking coefficient $S$ which remains uncertain. 
Studies of the sticking of H on grain surfaces
generally predicts that $S$ is about 1 at low temperatures and decreases with increasing temperature to about 0.4-0.1 at $T_{gas}\sim$300 K \cite[]{hollenbach79,burke83,leitch-devlin85,buch91}. Here, we approximate $S$ by 1/(1+ $T_{gas}$ / 400 K + ($T_{gas}$ / 400 K)$^2$) from \cite{burke83,bertoldi97}. 
Secondly, for the parameters of the physisorbed and chemisorbed
sites, we assume typical values 
for carbonaceous grains which are summarized in Table \ref{parameter}.
We treat the surface coverage of chemisorbed H atoms ($N_c/N$) as a free parameter. 
Further, we take $N_c/N$ for SGs and BGs to be equal and constant although this fraction is likely to depend upon grain characteristics and physical conditions.
This is not completely satisfactory but it gives some
 insight into this parameter which more detailed theories of these
 processes should attempt to fit.
We find that $N_c/N \sim$0.1 (see below)
 gives a reasonable fit to our PDR data.
Finally, we take the probability to eject H$_2$, $\eta$, to be unity.
This is based simply on the fact that ejecting the newly formed 
H$_{2}$ molecule \cite[the desorption energy is $\sim$0.05 eV,][]{katz99,cazaux2002a}
 requires a small fraction of the available formation
energy deposited in the grain (1.5 eV). Note that if we do not assume spontaneous desorption,
the H$_2$ formed would stay on the surface for really low temperatures (below 10 K)
 since the temperature is not high enough to allow evaporation \cite[see][]{cazaux2002,cazaux2002a}.

For the purpose of comparison with the formation rates
determined using PDR models (Sect. \ref{observations}), 
we need to estimate dust temperatures in the PDRs of interest to us.
In Table \ref{rate_formation}, we give the temperatures of SGs and BGs expected 
in the H$^0$/H$_2$ transition zone,
where the radiation field is given by $\chi~\exp{(-\tau)}$ with $\tau$ the FUV dust opacity in the transition zone equal to about unity.
For the big grains at thermal equilibrium, we have adopted
the analytic expression of \cite{hollenbach91}.
For the small grains subject to thermal fluctuations, 
we list in Table \ref{rate_formation} the temperatures for a graphite grain
of radius 1.5 nm ($N(C)\simeq 1600$). 
We have computed for each value of the radiation field
the full temperature histogram and
the values in Table \ref{rate_formation}
 correspond to the median value of the time dependent temperature (i.e., the grain spends half of the time at temperature higher and lower than this value).
 The median temperature of small grains is unlike that of big grains not constrained by their emission spectrum.
In these calculations, we used the absorption and emission cross section of bulk graphite from \cite{draine84}; the 
heat capacity is also of graphite given by \cite{guhathakurta89}.
We have thus ignored quantum effects which might significantly change properties
of these particles.
This estimate of the small grain temperature
 is only indicative. Ideally, one should 
take into account the time dependence of the temperature as a function of the grain
size. Further, their emission properties might not allow then to cool down to the 
cosmic 2.7 K background temperature.
Moreover, for very low temperatures ($\le 10$K), the heating due to the
collisions between the grains and warm gas ($T_{gas}\ge$100 K) 
or due to the formation of H$_2$ ($\sim$1/3 of the formation energy is deposited in the grain) would increase the grain temperature.
However, in the following we will use the SG temperatures
reported in Table \ref{rate_formation} in order to see 
if small grains are relevant for forming H$_2$.  
We find that the median temperatures of small grains are
significantly lower than those of big grains.
Indeed, SGs spend most of their time at low temperature
 \cite[]{guhathakurta89}.

\begin{table} \caption{Temperatures of dust and gas and H$_2$ formation rates}
\label{rate_formation}
\begin{tabular}{llllll}
\hline
Region & $T_{SGs}^a$ & $T_{BGs}^a$ & $T_{gas}^b$ & $R_f$$^c$\\
        & (K) & (K) & (K) & (cm$^{3}$ s$^{-1}$) \\
\hline
\hline
Chamaeleon   & $>$2.7  & 15 & 60 & 4 10$^{-17}$ \\
& & & \\  
Oph W         & 10  &36  & 330& 1.5 10$^{-16}$ \\
S140         & 10  &36  & 500& 1.5 10$^{-16}$ \\
IC 63        & 12  &44  & 620& 1.5 10$^{-16}$ \\
NGC 2023     & 25  &60  & 330& 3 10$^{-17}$ \\
Orion Bar    & 62  & 90 & 390& 3 10$^{-17}$ \\
\hline
\end{tabular}

$^a$ Temperature of the small grains (SGs) and big grains (BGs) expected in the 
H$^0$/H$_2$ transition zone (see text in Sect. \ref{comparison}).\\
$^b$ Gas temperature inferred from the distribution of low H$_2$ pure rotational levels (see Table \ref{pdrs}). For the Chamaeleon see \cite{gry2002}.\\ 
$^c$ H$_2$ formation rates derived from observations (see Sect. \ref{observations}). \\

\end{table}

\begin{figure*}[htbp]
\leavevmode
\begin{minipage}[c]{8.0cm}
\centerline{ \psfig{file=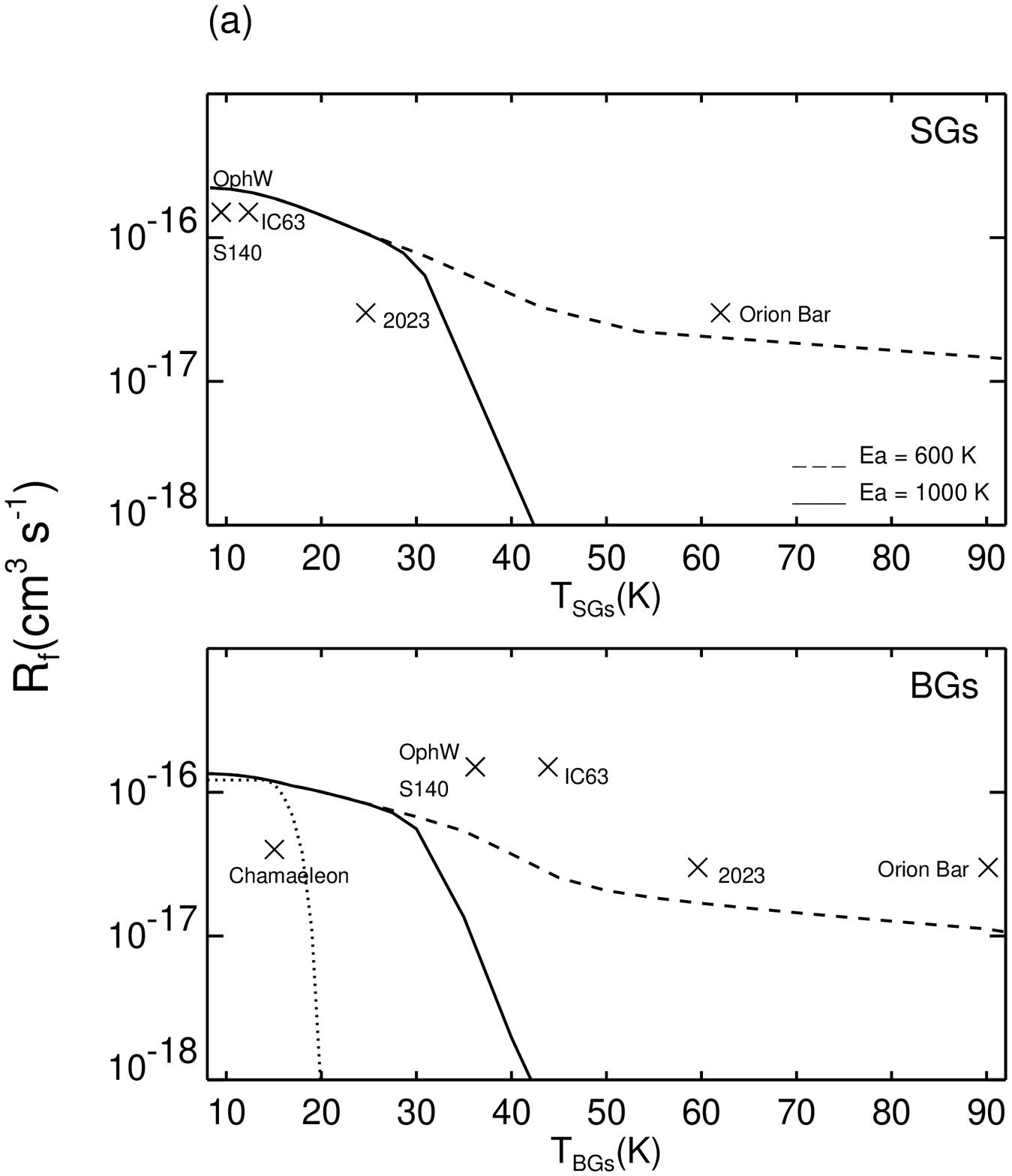,width=8.0cm,angle=0} }
\end{minipage}
\begin{minipage}[c]{8.0cm}
\centerline{ \psfig{file=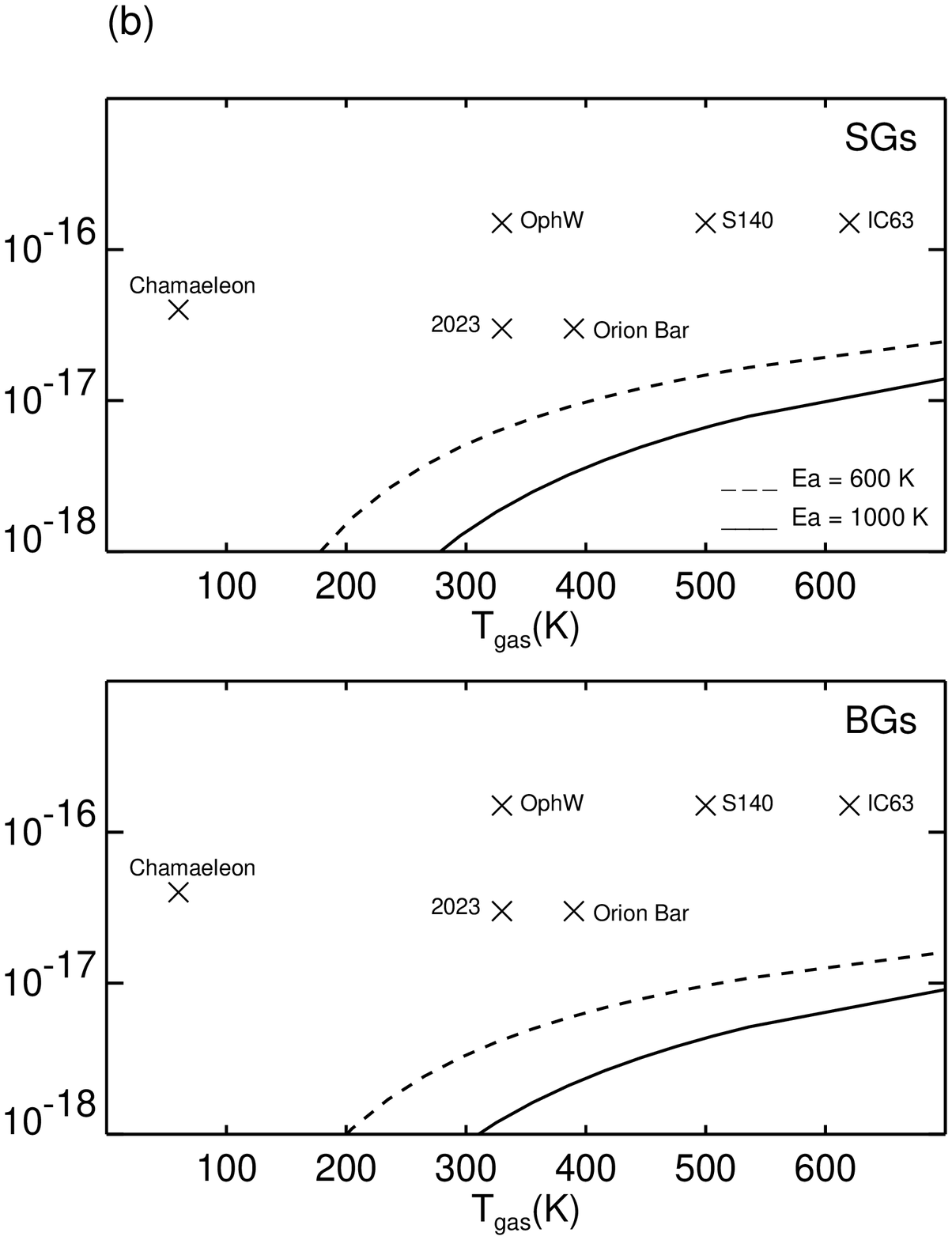,width=8.0cm,angle=0} }
\end{minipage}
\caption{\em Panel {\bf(a)} : H$_2$ formation rate as predicted for the indirect chemisorption mechanism
(solid and dashed lines) and for the pure Langmuir--Hinshelwood mechanism (dotted lines) and as 
observed (crosses, see Table \ref{rate_formation}) as a function of the 
dust temperature. The dust temperatures for Oph W
and S140 derived are similar (see Table \ref{rate_formation}).
The upper and lower panels show the rates predicted 
respectively for formation on small and big grains. 
The solid and dashed lines correspond respectively to $E_a$ equal to 1000 and 600 K.
Panel {\bf(b)} : H$_2$ formation 
rate as predicted for the H$_2$ formation by direct chemisorption (solid and dashed lines) and as observed (crosses) as a function of the
gas temperature. The ``observed'' gas temperatures are
 inferred from the distribution of low H$_2$ pure rotational levels reported in
Table \ref{pdrs}; for the Chamaeleon see \cite{gry2002}.}
\label{h2_formation}
\end{figure*}

 With the above formalism, we compute the formation rate $R_{f}$ for 
 the various processes outlined above on both SGs and BGs.  The
 results are presented in 
Fig. \ref{h2_formation}, where we compare the H$_2$ formation
rate  predicted for indirect and direct chemisorption with the rates derived from observations.
For completeness, we also show the results for the case
of the pure Langmuir--Hinshelwood mechanism upon BGs using the model 
of \cite{katz99} based on experimental results for graphite surfaces 
\cite[]{pirronello99a}.

First, we see from Fig. \ref{h2_formation}(a) that indirect chemisorption
may explain the data for all sources if the activation barrier energy
for recombination is $E_a \lesssim 1000$ K and if the fraction of chemisorbed H atoms
is as large we have assumed, $N_c/N\sim$0.1. This result is broadly  consistent with 
the recent theoretical model of \cite{cazaux2002,cazaux2002a},
 which accounts properly for the population of chemisorbed and physisorbed
 sites on the surface. Thus, our estimate of the fraction of chemisorbed H atoms
 may be reasonable. 
However, the value of the activation barrier energy is uncertain and critically influences our estimate for $R_{f}$ 
and the temperature range for which
  formation via indirect chemisorption is important.
We find in fact that for $E_a\sim$1000 K this process will be efficient until $T_{dust}\sim$30 K (i.e., important in PDRs of relatively moderate excitation, $\chi <$5000); 
for $T_{dust}>$30 K the physisorbed H atoms evaporate before recombining.
Conversely, for $E_a\sim$600 K the formation via indirect chemisorption 
will be still efficient at $T_{dust}\sim$100 K (i.e., important
even in highly excited PDRs).
In this last case, where the diffusion term dominates (see Sect. \ref{for_chem}), 
the decreasing of $R_f$ from moderate to highly excited PDRs
 is due to the competition between evaporation and finding
 the chemisorbed H atoms over the surface.


Secondly, we find that small grains which dominate
the total grain area and spend most of their time at low temperature ($<30$ K for $\chi \le$3000, see Table \ref{rate_formation}) 
may be the most promising surface for forming H$_2$ via indirect chemisorption.
The H$_2$ formation on SGs can be effective even in highly excited PDRs
 where the equilibrium temperature of BGs 
can be too high to form H$_2$.
In the case of SGs, we note that in most objects, the
observed H$_2$ formation rates can be explained with a high value
of $E_a$ ($\sim$1000 K). For BGs, conversely a low value of $E_a$
($\lesssim$600 K) is required.  
However, our description of the temperature of SGs
is crude (median value of the temperature distribution for a single 
size of grain). As a consequence, we cannot quantitatively constrain
the parameters ($E_a$, $E_p$, ..) of the H$_2$ formation process.

The direct or Eley--Rideal mechanism fails to explain the
  observations by a factor of a few (see Fig. \ref{h2_formation}(b)) although
 a higher $N_c/N$ combined with a slightly lower $E_a$ would suffice to explain some of the
  data. 
We cannot therefore exclude this mechanism but find that 
  our present data suggest that the indirect chemisorption on SGs is
  more probable. 

 There are several complications which we have neglected in the
  above treatment. One is the possible presence
  of interstitial sites 
as suggested by \cite{duley96} which can increase
the efficiency of  direct chemisorption. Another perhaps is 
 H$_2$ formation through a reaction between two 
chemisorbed H atoms  as suggested  by \cite{cazaux2002,cazaux2002a}. 
This mechanism would be efficient at high dust temperatures ($\gtrsim$100K) and could
be a possibility to explain the formation of H$_2$ in Orion.

 In summary, we conclude that in order to explain the H$_2$ formation efficiency
in PDRs,  the indirect chemisorption mechanism upon small grains is the most promising. 
This requires an activation barrier energy between a physisorbed
  H atom and a neighbouring chemisorbed H atom $E_a\lesssim$1000 K (or $\lesssim$0.1 eV) 
and a fraction of occupied chemisorbed sites of around ten percent.
 This conclusion is consistent
with our finding of a correlation between 
 the H$_2$ and PAH emission which 
suggests that $R_f$ scales with the PAH
abundance (see Sect. \ref{SGs_diagnostic}).
However, a better knowledge of the SGs properties (temperature, coverage of 
absorbed H atom, abundance ...) and of the mobilities of H atoms on realistic 
grain surfaces are critical.

\section{Conclusion}
\label{conclusion}

 The main aim of this study has been to provide estimates of the
 molecular hydrogen formation rate in a sample of nearby PDRs
 using results from both ISO and ground--based telescopes.  The
 physical conditions in the PDR layers from which H$_2$ emission is observed
 ($n_H\simeq 10^3-10^5$ cm$^{-3}$, $T_{gas}\ge$ 300 K)  differ considerably from those in the diffuse
 clouds where one can observe the UV lines of H$_2$ ($n_H \simeq$100-1000
 cm$^{-3}$, $T_{gas}\simeq$50-100 K) and upon which most estimates of the molecular
 hydrogen formation rate have been based.
 Thus  the results from PDRs allow important
 constraints to be placed upon the mechanisms for forming molecular hydrogen
 on grain surfaces. 
 We confirm the earlier result of \cite{habart2003a} that the H$_2$ formation
 rate in regions of moderate excitation ($\chi \le$1000) such as Oph W, S140 and IC 63
  is a factor
 of $\sim$5 times larger than the standard rate estimated by Jura for diffuse
 clouds (and confirmed by recent FUSE data). 
 On the other hand,  towards regions of higher radiation field
 such as the Orion Bar and NGC 2023, we derive H$_2$ formation rates consistent
 with the standard value.  Thus, the higher grain  and gas temperatures in
 PDRs do not seem to impede the formation of H$_2$.

  We have attempted to interpret these results with simple empirical models of the
  formation of H$_2$ on grain surfaces. From these, we conclude that an
  ``indirect chemisorption'' model analogous to that discussed by \cite{cazaux2002,cazaux2002a} is capable of explaining the data.
This result requires an activation barrier energy against the recombination of a physisorbed
  H atom and a neighbouring chemisorbed H atom $E_a\lesssim$0.1 eV.
Another condition appears to be that one needs an appreciable
fraction of surface sites occupied (few percent at least) with a binding energy of order 1 eV relative to the total number of surface binding sites (presumably mainly physisorbed with binding energies of order 0.05 eV).
Moreover, we suggest that small (size $<$ 100 \AA) grains 
may be the most promising surface for forming H$_2$ in PDRs. 
There is in fact enough grain surface in small grains to allow the formation
rate to be larger
than the standard H$_2$ formation rate and small grains spend most of their time at low temperature \cite[]{guhathakurta89}. H$_2$ formation by indirect chemisorption upon small grains should be effective even in highly excited
regions where large
grains are quite warm.

Our results show that formation of molecular hydrogen in PDRs is likely to take place with 
a different mechanism than in the diffuse interstellar medium
where the formation by physisorbed H atoms (Langmuir-Hinshelwood)
probably dominates.
In fact, in cold diffuse clouds the surface density of strongly bound H 
atoms should be low and consequently the formation by chemisorption would
 not be efficient.

   There are several fundamental uncertainties in  our present estimates
   of the H$_{2}$ formation rates in PDRs which future work should try to eliminate.
   One is due to the fact that we have used steady--state PDR models.
   This assumption can cause appreciable errors and it would be useful to calculate the expected H$_2$ line intensities
   for models with lower H$_{2}$ formation rates but where advection
   has been taken into account. 
However, \cite{stoerzer98} have modelled PDR structure assuming an ionization
front moving into the PDR and found that non-equilibrium effects are probably minor in objects similar to the Orion Bar.

   Other uncertainties are of an observational
   nature. It would be useful to have reliable estimates of the
   inclination and density distribution in the PDRs of interest
   in order to better constrain the models.
   There are indications in the
   Orion Bar for instance \cite[]{walmsley2000} that the column density is
   higher in the molecular layers of the PDR than in the region (discussed
   in this paper) where the H$_2$ lines are formed. Density gradients
   perpendicular to the PDR  photodissociation front clearly need to be
   taken into account when considering the spatial distribution of the
   various H$_2$ lines. 

   The theoretical models of H$_2$ formation discussed here are clearly very
   preliminary.   More detailed models need to explicitly consider the
   degree of occupation of chemisorbed sites by H-atoms
   as well as the mobility
   of H atoms on various types of grain surfaces. Our results however do
   suggest that more detailed consideration of H$_2$ formation on the
   surface of small grains would be worthwhile.  In this case,
   one should ideally follow the thermal fluctuations of these small particles
   and take into account the dependence on grain size in order to examine properly their contribution
to H$_2$ formation. Clearly
   also our present estimates for processes such as tunneling are very
   crude.

Additional information on the H$_2$ formation process could probably be obtained
   using high quality data for H$_2$ in excited vibrational states
which gives constraints on the excitation state of the newly formed H$_2$.
 It also would be useful to obtain estimates of the H$_2$ formation
 rate under a variety of different conditions. One such condition might be in
 the Magellanic clouds where the different metallicity, extinction curve,
 and radiation field  potentially may  influence both the available grain
 surface area and the efficiency of H$_2$ formation. Another is in the thin
 high excitation clouds   such as that found by \cite{meyer2001} towards
 the exciting star of NGC 2023.   It is clear that given our uncertainty
 about grain compositions and size distributions in different ISM
 locations, one is forced to some extent to use the astrophysical data to
 guide our estimates of processes such as H$_2$ formation.  This is perhaps
 philosophically not as satisfactory  as the traditional approach of
 employing experimentally determined or theoretically calculated rates but
 it is likely that nature does not give us the choice.

\acknowledgements{We are grateful to the referee, David Hollenbach, for relevant
comments and suggestions.
We also thank St\'ephanie Cazaux and Eric Herbst for fruitful discussion on the H$_2$
formation process. C. M. W. wish to acknowledge travel support from the MUIR project
``Dust and Molecules in Astrophysics Environments''.}

\newpage

{\large Appendix : Probability for the diffusion and recombination of
a physisorbed H atom with a chemisorbed H atom}
\par\bigskip\noindent
\par\bigskip\noindent
To calculate the probability $f=\frac{\tau _{p}^{-1}}{\tau _{p}^{-1}+\tau _{ev}^{-1}}$
(Eq. \ref{Eq:f_ind_che}, see Sect. \ref{for_chem}), we estimate 
the timescale $\tau _{p}$ for diffusion
over the grain surface followed by recombination and the timescale
 $\tau _{ev}$ for evaporation
 of the physisorbed H.
The latter time scale is given by 
\begin{equation}
\tau _{ev} = \nu _0^{-1} \exp{\left(\frac{E_d}{kT_{dust}}\right)}
\end{equation}
with $E_d$ the desorption energy
of a physisorbed H and $\nu _0$ the vibrational frequency of H in a physisorbed site typically of the
order of 10$^{12}$ s$^{-1}$. $\tau _{p}$ can be given by the sum of the time to 
find a neighboring site to a chemisorption site bearing
a H atom, $\tau _{m}$ 
and of the time to recombine and form H$_2$, $\tau _{rec}$.
We assume that diffusion from one physisorbed site to another 
occurs by thermal hopping \cite[]{katz99}. The time to hop
from one physisorbed site to the next is thus given by $\nu _0^{-1} \exp{\left(\frac{E_p}{kT_{dust}}\right)}$ with $E_p$ the activation barrier energy for physisorbed H atom diffusion.
The mobility is a random walk and considering that there are 4 neighboring sites 
for each chemisorbed site, we find that one needs approximately $(N/4N_c)^2$ steps to be adjacent to a filled chemisorbed site.
Thus:
\begin{equation}
\tau _m=\left(\frac{N}{4N_c}\right)^2 \times \nu _0^{-1}  \exp{\left(\frac{E_p}{kT_{dust}}\right)}
\end{equation}

The physisorbed H atom must cross the activation barrier 
to recombine with the nearest chemisorbed H atom by either (i)
thermal diffusion with a probability $f_{th}=\exp{(-\frac{E_a}{kT_{dust}})}$ or by (ii) tunnelling with a probability
 $f_{tun}=\exp{\left(-\frac{2~\Delta x}{\hbar}~(2 m_H E_a)^{0.5}\right)}$ \cite[]{messiah72} with $\Delta x$ the width
of the barrier.  
The recombination
time scale can be approximated by 
\begin{equation}
\tau _{rec} = \left(\frac{N}{4N_c}\right) \times \nu _0^{-1}   \frac{1}{(f_{th}+f_{tun})}
\end{equation}
At high dust temperatures (above $\sim$40 K) thermal
   diffusion dominates whereas at low temperature (below $\sim$30 K) tunneling is more important. 

There are many caveats to the above procedure.
In particular, we note that  
considering the probability $\exp{\left(-\frac{2~\Delta x}{\hbar}~(2 m_H E_a)^{0.5}\right)}$ for 
tunneling transmission through a single barrier is incorrect for relevant 
astrophysical surfaces which are not regular and need therefore to be 
modeled by a set of barriers as many as there are atoms. The 
tunneling probability we adopt is thus an upper limit. However, for  
the purpose of comparison 
with observationally determined rates, our approach allows us to see 
which of the H$_2$ formation processes considered in this study will dominate. 

\newpage

\bibliographystyle{natbib}

\end{document}